\documentclass[prb,twocolumn,showpacs]{revtex4}
\usepackage{graphics}
\begin{document}
\title{Colossal enhancement of magnetoresistance in
La$_{0.67}$Sr$_{0.33}$MnO$_{3}$ / Pr$_{0.67}$Ca$_{0.33}$MnO$_{3}$ multilayers:
reproducing the phase-separation scenario}
\author{Soumik Mukhopadhyay
\email{\sf{soumik.mukhopadhyay@saha.ac.in}}}
\author{I. Das}
\affiliation{ECMP Division, Saha Institute of Nuclear Physics, 1/AF,
Bidhannagar, Kolkata 700 064}

\begin{abstract}
Colossal enhancement of magnetoresistance has been achieved
over a broad temperature range which extends upto the room temperature, 
in ferromagnetic
metal-charge ordered insulator manganite multi-layers. 
The artificially created
phase coexistence in the multilayers reproduce the characteristic
signatures of metastability in the magnetotransport properties  
commonly observed in electronically phase-separated manganites. 
\end{abstract}


\maketitle

There are numerous experimental evidences available
now which confirm that transition
metal oxides such as manganites are electronically 
inhomogeneous~\cite{uehara, fath, zhang, loudon, ddsarma}.
The length scale of these inhomogeneities or the so called
phase separation varies from nanometer up to few microns.
Moreover, these inhomogeneities can be manipulated by application
of external stimuli such as temperature, magnetic field, etc.
Theoretically, however, only nanoscale inhomogeneities have been
predicted. It is generally believed that only mesoscale inhomogeneities
can give rise to colossal magnetoresistance. Uehara et al.~\cite{uehara}
explained the colossal magnetoresistance in a system of
coexisting phases such a ferromagnetic metallic (FM) and antiferromagnetic
insulator (AFI). The spin alignment among the FM domains are random
which becomes aligned on application of magnetic field resulting in
gigantic enhancement of conductivity. It was suggested that such
mesoscopic phase separation was necessary for colossal responses.
Besides this spatially static phase-separation picture, Fath et al.~\cite{fath}
presented an alternative explanation where the increase of FM volume
fraction at the cost of AFI domain on application of magnetic field
forming a percolation cluster leads to the colossal enhancement
of conductivity. Later on, different experimental techniques such as
magnetic force microscopy, transmission electron microscopy, 
photoemission spectromicroscopy have been employed for probing
phase separation~\cite{zhang, loudon, ddsarma}. Zhang et al.~\cite{zhang} showed
that magnetic domains evolve with temperature resulting in magnetic hysteresis
which coincides with the hysteresis in resistivity. Loudon et al.~\cite{loudon}
suggested that the mesoscopic ferromagnetic region is itself inhomogeneous
at the nanoscale with coexisting metallic and charge-ordered regions.
Sarma et al.~\cite{ddsarma} indicated a possible memory effect 
associated with the
electronic inhomogeneities. 
Apart from the microscopic probes, 
the electronic phase separation can be detected
by studying the macroscopic properties such as
transport and magnetization. Very recently,
magnetic structure of a series of
La$_{0.67}$Sr$_{0.33}$MnO$_{3}$ (LSMO) / Pr$_{0.67}$Ca$_{0.33}$MnO$_{3}$ (PCMO) 
superlattices has been studied,
and the possibility of tuning phase separation by
imposing appropriate geometrical constraints which
favor the accommodation of FM nanoclusters within the ``non-FM'' material
has been emphasized~\cite{fmafm6}.
In this article, we will show that if the thickness of the alternate
layers in LSMO/PCMO superlattice is properly tuned,
it is possible to mimic the characteristic transport and magnetic
responses of spontaneously phase separated manganites.

\begin{figure}
\resizebox{8.5cm}{7.5cm}
{\includegraphics{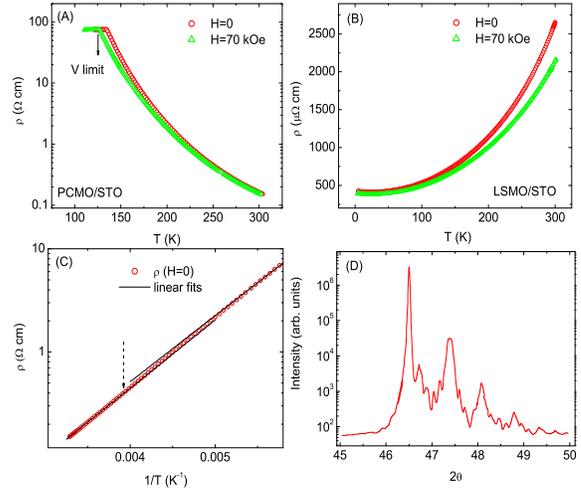}}
\caption{(Color online)
The temperature dependence of resistivity in absence as well as in
presence of strong magnetic field for A) PCMO and B) LSMO films.
C) Logarithm of resistivity plotted against inverse of temperature
for PCMO. The linear fits indicate two different regions. The blown out
portion of the transition region is shown
where the arrow-head corresponds to $T_{co}$. D) A representative
x-ray diffraction $\theta-2\theta$ scan for one of the multilayers (LSPC3)
showing oscillation of the Bragg-peak (002) characteristic of superlattice
structures.}
\label{fig:refer}
\end{figure}

Manganite multilayers are also interesting from the technological 
point of view since they are capable of
exhibiting colossal magnetoresistance.
The systems which exhibit large magnetoresistance (MR) are technologically
useful for application as magnetic field sensors and memory devices.
Microfabricated devices such as magnetic tunnel 
junctions~\cite{moodera, soumik1} 
do exhibit large MR. However, the sample fabrication
is extremely difficult, costly and time-consuming. Moreover, there are
other serious issues such as the sharp
bias dependence of tunnel magnetoresistance (TMR)~\cite{moodera, soumik2} 
and enhanced noise due to
high junction resistance~\cite{noise1} or magnetic fluctuations~\cite{noise2}. 
In this respect, manganite multilayers can become a useful alternative
in that they can be easily grown and do not necessarily need
to be microfabricated.
The resistance is much less compared
to the MTJs and hence the noise level is low. 
Very recently, it was shown 
that the reduction in individual layer thickness in an all-ferromagnetic
manganite multilayers can lead to giant enhancement of room temperature
MR~\cite{soumik3}, 
followed by observation of large room temperature MR in 
another all-ferromagnetic manganite superlattice~\cite{adv}.
On the other hand, there are several reports on 
ferromagnetic/ antiferromagnetic manganite superlattice
systems~\cite{fmafm6,fmafm1,fmafm2,fmafm3,fmafm4,fmafm5,fmafm7}.
However, in all those cases, as will be discussed later on, 
the magnetoresistive response is either miniscule
or confined within a very short temperature range away 
from room temperature.  
In this letter, we will also show that 
it is possible to achieve colossal magnetoresistive response
over a wide temperature range even
in ferromagnetic metal - antiferromagnetic charge-ordered insulator 
LSMO/PCMO
multilayers. 

\begin{figure}
\resizebox{8.5cm}{7.5cm}
{\includegraphics{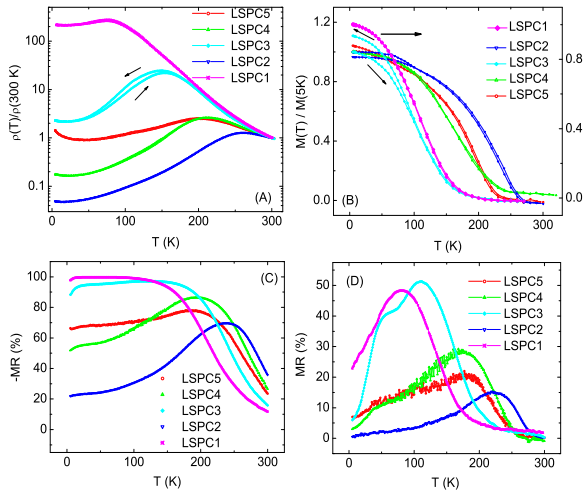}}
\caption{(Color online) A: The temperature dependence of zero-field
cooled resistivity during the cooling and warming cycle for all the samples.
B: Field cooled magnetization vs.
temperature curves for all the samples with data taken in the cooling
and warming cycles, LSPC3 showing significantly large hysteresis. The
hysteresis loop is open since here the warming cycle is followed by the
cooling cycle.
C: The temperature dependence of magnetoresistance at $70$ kOe for all
the samples. D: The same at $5$ kOe. The magnetoresistance is defined
as $\Delta\rho/\rho_{0}=(\rho_{H}-\rho_{0})/\rho_{0}$
}\label{fig:res}
\end{figure}

Five multilayers have been prepared using pulsed laser ablation of 
La$_{0.67}$Sr$_{0.33}$MnO$_{3}$ (LSMO) and Pr$_{0.67}$Ca$_{0.33}$MnO$_{3}$
(PCMO) ceramic targets, deposited on SrTiO$_{3}$
substrates held at a temperature $800^{\circ}$C in 350 mTorr oxygen pressure. 
The prepared multilayers can be divided into two sets having different
volume ratio of LSMO and PCMO. The first set contains multilayers
where the volume ratio of LSMO and PCMO is 1:1, while for the second set
the volume ratio is 3:2. The detailed description of the different samples
has been given in Table:~\ref{t.table}. The x-ray diffraction $\theta-2\theta$
scan plotted in Fig:~\ref{fig:refer}D shows the ($002$)  
distinct superlattice reflections (Fig:~\ref{fig:refer}D) 
for LSPC3. The sample diffraction peaks are as sharp as the substrate
peak, indicating high structural quality.
The calculated
out-of-plane lattice parameters for LSMO and
PCMO for the multilayers turn out to be $3.85$ and $3.75 \AA$, 
respectively.
These values being smaller than
the corresponding bulk values 
$3.88$ and $3.83 \AA$, respectively, confirms
the highly epitaxial nature of the multilayers and the bi-axial 
in-plane tensile stress due to the substrate. 
The magnetotransport properties
have been studied using the usual four probe method with the magnetic
field applied parallel to the electric field.  
\begin{table*}
\caption{Detailed description of the multilayers studied
}\label{t.table}
\begin{center}
\begin{tabular}{|l|l|l|l|l|l|l|l|l|l|} \hline\hline
Sample & Description & $\rho_{5K}$ & $\rho_{300K}$ & $T_{C}$ & $T_{MI}$\\
name &  & ($\Omega cm$) & ($\Omega cm$) & (K) & (K) \\ \hline
LSPC1 & LSMO$_{10\mathrm{\AA}}$/PCMO$_{10\mathrm{\AA}}$ & $35.2$ & $0.09$ & $110
$ & $75$  \\
LSPC2 & LSMO$_{15\mathrm{\AA}}$/PCMO$_{10\mathrm{\AA}}$ & $0.004$ & $0.07$ & $24
0$ & $262$ \\
LSPC3 & LSMO$_{15\mathrm{\AA}}$/PCMO$_{15\mathrm{\AA}}$ & $0.22$ & $0.1$ & $100$
 & $150$ \\
LSPC4 & LSMO$_{22.5\mathrm{\AA}}$/PCMO$_{15\mathrm{\AA}}$ & $0.01$ & $0.06$ & $1
60$ & $213$ \\
LSPC5 & LSMO$_{22.5\mathrm{\AA}}$/PCMO$_{22.5\mathrm{\AA}}$ & $0.13$ & $0.1$ & $
200$ & $200$ \\ \hline\hline
\end{tabular}
\end{center}
\end{table*}

In addition, as reference samples, transport properties of 
LSMO and PCMO films of thickness 
$500\mathrm{\AA}$ deposited on STO substrates have been studied. 
The resistivities of both films show very weak sensitivity to magnetic 
field (Fig:~\ref{fig:refer}A, B). It is obvious
from Fig:~\ref{fig:refer}A 
that even a magnetic field as high as $70$ kOe is unable to destabilize, 
presumably, what should essentially be the 
charge-ordered state in PCMO.
It is difficult to pinpoint the charge ordering transition
from magnetization measurement due to extremely weak magnetic signal
of the film superimposed on the large diamagnetic signal of the substrate.
The charge ordering temperature can be ascertained
from the temperature dependence of resistivity using the methodology
employed in ref~\cite{epl}, in which 
$T_{co}$ can
be determined from the plot of $\log\rho$ against $1/T$, as
shown in Fig.~\ref{fig:refer}C. The plot shows three regimes where the high
temperature and the low temperature regime is separated by 
a transition region within which $\rho$ increases more
strongly. In this case, $T_{co}$ turns out to be $255$ K, well above
the $T_{co} = 220 K$ for the corresponding bulk sample~\cite{anis}.
This is consistent with previous results where enhancement of $T_{co}$
due to bi-axial strain (compared to the bulk) was observed in films  
~\cite{apl}. The critical magnetic field for charge-order
melting in the corresponding bulk sample is about $60$ kOe~\cite{tomioka} 
while for the
present case, the critical field seems to be beyond $70$ kOe which is
consistent with the enhanced value of $T_{co}$. 
The biaxial strain can influence two important parameters
which determine the co-operative interaction and hence the transport
or the magnetic properties: $1)$ The extension or
contraction of the Mn-O-Mn bond-length leads to a large reduction or
enhancement of the electronic hopping amplitude;
$2)$ the
increased Jahn-Teller distortion leads to localization of
electrons. The experimental results for PCMO film described here 
are in agreement with already
published report~\cite{apl} and suggest that the bi-axial
strain imposed by the substrate stabilizes the charge ordering in 
PCMO film.

\begin{figure*}
\resizebox{16cm}{9cm}
{\includegraphics{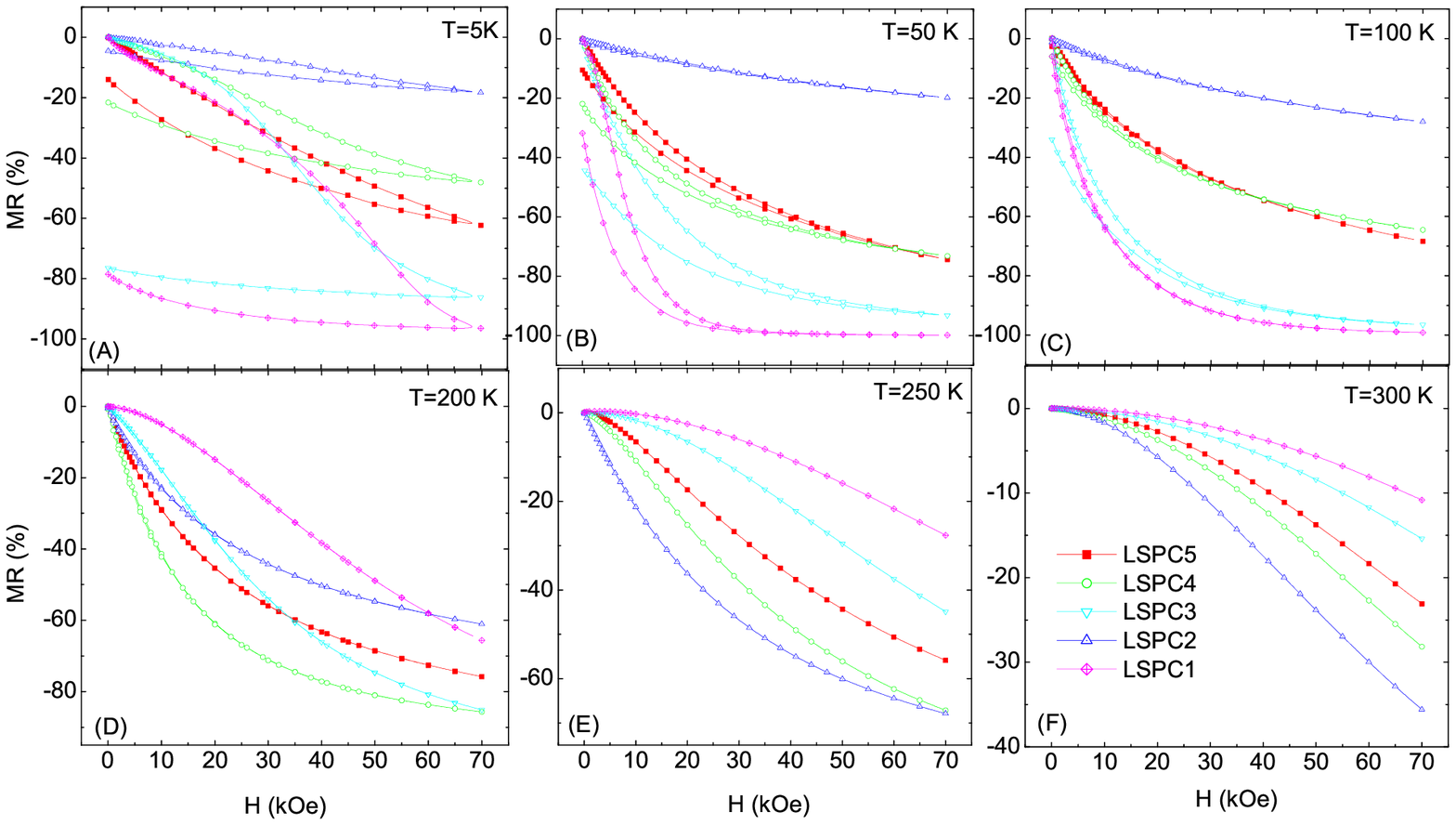}}
\caption{(Color online)
The magnetic field dependence of magnetoresistance for all the samples
at different temperatures. All the multilayers exhibit varying degree
of metastability at low temperature.
}\label{fig:hyst}
\end{figure*}
 
The resistivity and the field cooled magnetization 
of all the samples in absence of magnetic field during
the cooling and warming cycle has been plotted in Fig:~\ref{fig:res}A, B
respectively.
There are a few interesting points  
$1)$ Keeping the LSMO/PCMO volume ratio constant at $1:1$, as the number
of interfaces is increased, the low temperature resistivity increases
significantly. 
$2)$ As the LSMO volume fraction is increased the temperature dependence
of resistivity becomes distinctly metallic with high value of $T_{MI}$.
Keeping the volume ratio constant, the increase in PCMO layer 
thickness leads to reduction in $T_{MI}$ value. $3)$ Particularly
for LSPC3, both the resistivity and the field-cooled magnetization
curves during the cooling and warming cycle shows hysteresis.

Although the parent materials LSMO and PCMO hardly exhibit 
any significant MR within the temperature range $5-300$ K,
extra-ordinary enhancement of MR has been observed 
for all the multilayers (Fig:~\ref{fig:res}C, D and
Fig:~\ref{fig:hyst}). 
LSPC1 and LSPC2 exhibits close to $100\%$ MR (here the MR has been
calculated using the so-called pessimistic definition) over a wide
temperature range and the magnetoresistance is quite large
even at low magnetic field (Fig:~\ref{fig:res}D). 
As the thickness of the LSMO layer 
or the volume fraction of LSMO is further increased, 
large magnetoresistance
is observed near room temperature. 

The electronic conduction
in manganites is governed by the competition between different 
energy scales such as the kinetic energy of the $e_{g}$ electrons and the
on-site Coulomb repulsion. If both energy scales become comparable,
which is most likely for narrow bandwidth systems, there is a tendency
towards phase separation into ferromagnetic metallic and anti-ferromagnetic
charge-ordered insulating regions~\cite{amlan}.
In this case, the phase coexistence has been introduced artificially 
by depositing
alternate LSMO and PCMO layers of nanoscale thickness, and the 
magnetotransport properties observed mimic the properties exhibited
by electronically phase-separated manganites, 
for example, metastability, long-time
magnetic relaxation~\cite{amlan}, etc. 

\begin{figure}
\resizebox{8.5cm}{7.5cm}
{\includegraphics{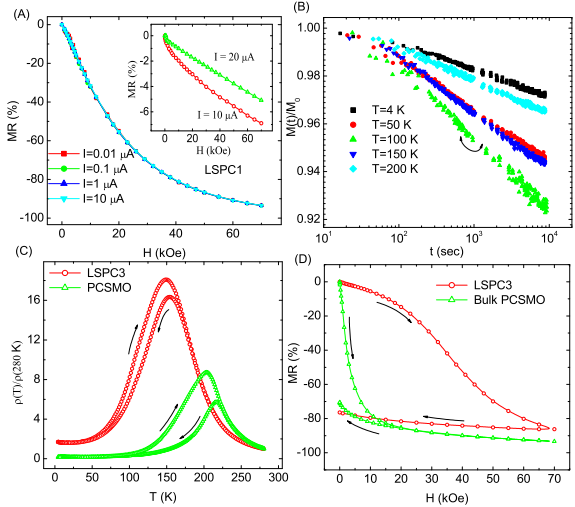}}
\caption{(Color online) A) The bias dependence of MR (the bias dependence
of MR of a granular NSMO film is shown in the inset for comparison)
at $200$ K for LSPC1. B) The magnetic relaxation data for LSPC3
at different temperatures
C) The temperature dependence of resistivity and D) the
magnetic field dependence of resistivity for LSPC3 and bulk PCSMO (for
comparison) at $4$ K
showing signatures
of metastability.
}\label{fig:mrbias}
\end{figure}

Fig:~\ref{fig:hyst}A gives strong
evidence of metastability at low temperature in that the low-resistance-state
obtained after the sample is exposed to strong magnetic field is retained even
after the magnetic field is removed. The high resistance state is recovered
when the temperature is raised to sufficiently high value and then zero-field
cooled back to the low temperature.  
As the temperature is increased,
due to thermal fluctuation, the width of the hysteresis curve gets reduced.
The rate at which the magnetic field is swept, has been kept constant
for all measurements.
The temperature and magnetic field dependence of resistivity of a typical 
electronically phase-separated polycrystalline manganite sample 
Pr$_{0.65}$(Ca$_{0.6}$Sr$_{0.4}$)$_{0.35}$MnO$_{3}$ (PCSMO) has been compared
with that for LSPC3 (Fig:~\ref{fig:mrbias}D), 
showing similar evidence of metastability. 
This metastability has been
directly verified from magnetic relaxation measurements
carried out at different constant temperatures 
over the temperature range $4-250$ K.
A magnetic field of $50$ kOe was applied for $60$ sec after which the
field was removed, followed by data collection over a time-span of 2 hours.
The magnetic relaxation data shown in Fig:~\ref{fig:mrbias}B
corresponds to LSPC3.
The long-time relaxation is clearly logarithmic and the negative slope-value
of the reduced magnetization vs. time curves increases systematically
as the temperature is decreased up to $100$ K. Below $100$ K, a completely
opposite trend is observed (Fig:~\ref{fig:mrbias}B).
The magnetic relaxation at long time scales can be described
approximately by the expression $M(t)/M(t_{n})=1+S\log(t/t_{n})$.
Here, $S$ is called magnetic viscosity, and $t_{n}$ and $M(t_{n})$ are
the normalization time and the corresponding magnetization
at that point in time, respectively.
The logarithmic relaxation can be attributed
to a free energy landscape containing local minima
corresponding to different equilibrium states separated
by energy barriers. 
Or to put it in another way, after the magnetic field is removed, the
charge-ordered fraction tries to reappear at the expense of ferromagnetic
fraction, resulting in long-time relaxation.
At low
temperature, after removing the magnetic field, the system gets trapped
in a frozen cluster-glass like state due to the frustration at
ferromagnetic (FM)/ antiferromagnetic (AFM) interfaces which relaxes 
extremely slowly with time, compared to higher intermediate 
temperature. 
Such long-time relaxation is only observed for LSPC1 and LSPC3, the
rest of the samples do not exhibit any significant magnetic relaxation.
Interestingly, so far as magnetotransport is concerned, the manifestation of 
metastability is more distinct in LSPC1 and LSPC3, compared to the other
samples. As the thickness of the LSMO and PCMO layers is increased 
the observed metastability is comparatively less pronounced, 
suggesting the importance of the length scale of these 
inhomogeneities. Moreover, if the volume fraction of LSMO is increased
compared to PCMO, the signatures of metastability get even weaker.

The highest value of MR at room temperature is about $36\%$ 
(Fig:~\ref{fig:hyst}F),
which is comparable to or better than that corresponding to all-ferromagnetic
manganite multilayers reported so far~\cite{soumik3, adv}.
The direct exchange coupling between FM/AFM layers due to the
proximity of the ferromagnetic LSMO
layers can induce an effective internal field~\cite{fmafm1, inter2}
The large value of MR observed in our case can be attributed to the
lowering of critical magnetic field needed externally 
for melting of charge ordered state
due to the introduction of ferromagnetic layers which can produce
large internal magnetic field.
In bulk PCMO samples,
the ground state is a charge-ordered (CO)
AFM phase~\cite{tomioka}. However, for LSMO/PCMO or LCMO/PCMO multilayers, 
it has been observed that while the PCMO layers are
still AFM, the CO state is not realized~\cite{fmafm6, fmafm2, fmafm7}. Due
to the strong electron-phonon coupling, strain fields play
a key role in the formation of CO phase. Thus the suppression of
charge ordering in the above-mentioned experiments 
must be related with lattice strain and interface clamping~\cite{fmafm6}.
Even in our case, since the critical magnetic field for melting of 
charge ordering in the PCMO film is distinctly higher compared to the
bulk, the effect of 
strain fields cannot be ruled out.

All the multilayers show remarkable bias-stability at all temperature,
over at least $4$ orders of magnitude (Fig:~\ref{fig:mrbias}A, B).
This suggests that the
multilayers are free from extrinsic effects such 
as spin dependent inter-granular 
transport~\cite{soumik4, soumik5}.
The sharp bias dependence of MR for a granular 
Nd$_{0.67}$Sr$_{0.33}$MnO$_{3}$ is shown in the
inset of Fig:~\ref{fig:mrbias}A, for comparison. 
The negligible bias dependence of MR is ideally suited
for application as magnetoresistive sensors. The fact that there is no
hysteresis in the magnetic field dependence of MR 
above $100$ K, adds to the prospect of the multilayers being qualified
as potential magnetoresistive sensors. The resistivities of all the samples
are reasonably low and epitaxial manganite films are known to
exhibit extremely low $1/f$ noise~\cite{noise}.  

As mentioned in the introduction, there are a few
reports on all-manganite FM-metal/AFM-insulator 
superlattice systems~\cite{fmafm6, fmafm1, fmafm2, fmafm3, fmafm4, 
fmafm5, fmafm7}. 
While
some of them deals with magnetic properties and 
exchange biasing effect at the 
FM-AFM interfaces~\cite{fmafm1, fmafm3, fmafm5, fmafm6},
in some cases, magnetotransport properties have also been studied. 
For example, large low field magnetoresistance (LFMR) has been observed in
LCMO/PCMO superlattice~\cite{fmafm2,fmafm7}. 
However, in such cases there is hardly
any evidence of the influence of 
charge ordering and the LFMR is observed round about
the Curie temperature of LCMO and the enhancement of MR compared to
the parent LCMO film (which is known to exhibit colossal MR
anyway) is not substantially high. In our case, the enhancement of 
MR has been observed over a large temperature range,
without using colossal magnetoresistive material such as LCMO.
Moreover, the magnetotransport properties give strong evidence of 
the influence of charge ordering, which can be easily destabilized using
magnetic field, in complete contrast to the highly stable charge-ordered
state in PCMO film described here and reported by other group~\cite{apl}.

To conclude, colossal enhancement of MR has been
achieved over a wide temperature range in LSMO/PCMO multilayers,
rarely observed in any manganite multilayers till date. 
The colossal magnetoresistance has been attributed to the melting
of charge-ordered state by magnetic field.
There
is no fundamental difference between the 
observed transport and magnetic properties of the multilayers 
(such as metastability and long-time relaxation) where the
nanoscale phase separation has been introduced artificially and that
of spontaneously phase-separated manganites.

\end{document}